\documentclass[12pt]{article}
\usepackage{graphicx,amsmath}
\usepackage{units}

\newlength{\dinwidth}
\newlength{\dinmargin}
\renewcommand{\vec}[1]{\boldsymbol{#1}}

\def\lapproxeq{\lower .7ex\hbox{$\;\stackrel{\textstyle                                                    
<}{\sim}\;$}}                                                    
\def\gapproxeq{\lower .7ex\hbox{$\;\stackrel{\textstyle                                                    
>}{\sim}\;$}}                                                    
\def\be{\begin{equation}}                                                    
\def\ee{\end{equation}}                                                    
\def\bea{\begin{eqnarray}}                                                    
\def\eea{\end{eqnarray}}

\def\sh{\hat s}
\def\sh2{{\hat s}^2}

\begin{document}                                                    
\titlepage                                                    
\begin{flushright}                                                    
IPPP/11/30  \\
DCPT/11/60 \\                                                    
\today \\                                                    
\end{flushright} 
\vspace*{0.5cm}
\begin{center}                                                    
{\Large \bf Probes of multiparticle production at the LHC}\\

\vspace*{1cm}
                                                   
M.G. Ryskin$^{a,b}$, A.D. Martin$^a$ and V.A. Khoze$^{a,b}$ \\                                                    
                                                   
\vspace*{0.5cm}                                                    
$^a$ Institute for Particle Physics Phenomenology, University of Durham, Durham, DH1 3LE \\                                                   
$^b$ Petersburg Nuclear Physics Institute, Gatchina, St.~Petersburg, 188300, Russia

\vspace*{1cm}                                                    
                                                    
\begin{abstract}                                                    
We discuss how the main features of high-energy `soft' and `semihard' $pp$ collisions may be described in terms of parton cascades and multi-Pomeron exchange. The interaction between Pomerons produces an effective infrared cutoff, $k_{\rm sat}$, by the absorption of low $k_t$ partons. This provides the possibility of extending the parton approach, used for `hard' processes, to also describe high-energy soft and semihard interactions. In particular, the presence of the cutoff
 $k_{\rm sat}$, which increases with collider energy, means that the production and hadronization of minijets with 
$p_t\gapproxeq k_{\rm sat}$ are the main source of the soft secondaries. We propose several measurements which can be made at the LHC, that can further illuminate our understanding of the mechanism which drives the soft and semihard interactions. We show that the structure of Pomeron-Pomeron interactions may be studied in Double-Pomeron-Exchange processes and in the $p_t$ distributions of secondaries near the edge of large rapidity gaps.  The so-called `ridge effect' is another manifestation of the interaction of two partonic cascades.  

\end{abstract}                                                        
\vspace*{0.5cm}                                                    
                                                    
\end{center}                                                    
                                                    
\section {Introduction}

`Soft' and `hard' high-energy hadron interactions are usually described in different ways.  The appropriate formalism for high-energy soft interactions is based on Reggeon Field Theory
\cite{RFT,ABK} with a phenomenological (soft) Pomeron, whereas for hard interactions we use QCD and a partonic approach. However, the two approaches appear to merge naturally into one another.  That is, the partonic approach seems to extend smoothly into the soft domain. The success of this picture is demonstrated in \cite{KMRnnn}.
Moreover, there are phenomenological arguments (such as the small slope of the Pomeron trajectory, the success of the additive quark model relations, etc.) which indicate that the size of an individual Pomeron is relatively small as compared to the size of a proton or pion etc. Thus it is natural to describe the Pomeron in terms of QCD, where it is associated with the BFKL vacuum singularity\footnote{That is, by the leading Regge $t$-channel exchange trajectory with vacuum quantum numbers.}\cite{BFKL}.

The BFKL equation describes the development (evolution) of the gluon/parton shower as the momentum fraction, $x$, of the proton carried by the parton decreases.  That is, the evolution parameter is ln$(1/x)$, rather than the ln$k_t^2$ evolution of the DGLAP equation. Actually, both evolutions have a common origin; they  result from evolution which is strongly ordered in the angles of the emitted partons. In DGLAP collinear evolution the angle $(\theta \simeq k_t/k_\ell)$ increases due to the growth of $k_t$, while in BFKL the angle grows due to the decreasing longitudinal momentum fraction
($k_\ell=xp$) as we proceed along the emission chain from the proton.  When we account for higher orders\footnote{Each extra order 
 allows that one 
additional adjacent parton along the chain ceases to be strongly ordered.} (NLO, NNLO,...), DGLAP evolution includes further elements of BFKL evolution, and vice-versa.

Formally, to justify the use of perturbative QCD, the BFKL equation should be written for gluons with large $k_t$. However, it turns out that, after accounting for NLO corrections and performing an all-order resummation of the main higher-order contributions, the intercept of the BFKL Pomeron depends only weakly on the scale. The intercept is found to be $\Delta \equiv \alpha_P(0)-1 \sim 0.3$ over a large interval of $k_t$
{\cite{bfklresum,kmrsre}.
Thus the BFKL Pomeron is a natural object to continue from the `hard' domain into the `soft' region.

\section{The `BFKL' Pomeron}

At this stage it is worth summarising what we mean by the BFKL Pomeron. Let us trace its origin from the beginning. The very early high-energy hadronic data appeared to be driven by a flavour-independent Regge trajectory with intercept $\alpha(0) \simeq 1$. A physical picture of how this object may be related to QCD was first given by Low and Nussinov \cite{LN}. They proposed that high-energy elastic scattering should be
 described by the two-gluon exchange. Since the gluon has spin 1, the 
 single-gluon exchange amplitude $\sim s$ and the two-gluon exchange amplitude $\sim (i/s)ss=is$ (where $i/s$ arises from integration over the gluon loop) which is just the phase and energy dependence of an even-signature trajectory with $\alpha=1$. The infrared divergences of massless gluon exchange cancel for the interaction of two colourless composite hadrons. At high-energies there is a huge probability to emit additional gluons so the Low-Nussinov $|M(2\to 2)|^2$ Pomeron becomes the $|M(2 \to n)|^2$ ladder structure (see Fig. \ref{fig:P}), corresponding to a trajectory $\alpha_P(t)$ with $\alpha_P(0) > 1$.  The next development was to allow for the interaction between neighbouring gluons. The NLO contributions and all-order resummation then results in a compact object which, here, we call the QCD BFKL Pomeron, which includes a mixture of BFKL together with some DGLAP properties. Such a formalism describes heavy onium-onium \cite{BL} scattering where the quark mass provides the large scale. For smaller scales  the QCD coupling, $\alpha_s$, increases. Moreover,  accounting for $2 \to n$ particle production, the amplitude of Pomeron exchange grows with energy. Thus we have to consider the exchange of many Pomerons. Then, with this increased number of Pomerons, we must allow for the possible interactions between them. That is, we must consider complicated multi-Pomeron diagrams.
\begin{figure} 
\begin{center}
\includegraphics[height=4cm]{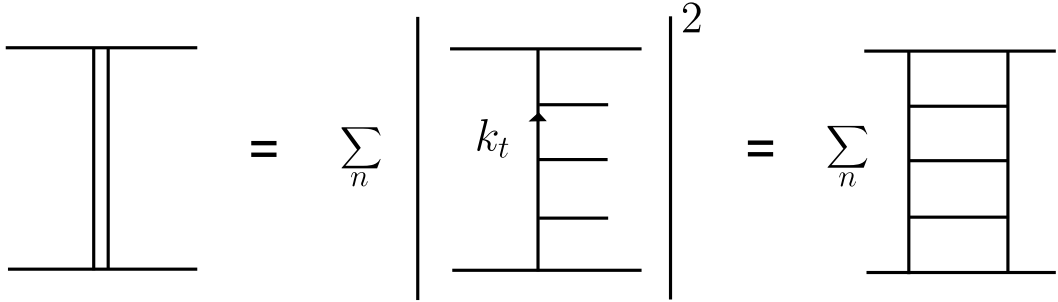}
\caption{\sf The ladder structure of Pomeron exchange.}
\label{fig:P}
\end{center}
\end{figure}

\section{Multi-Pomeron `BFKL' description of `soft' data}

Now we come to the discussion of the LHC `soft' data. Here the interval of BFKL ln$(1/x)$ evolution is much larger than that for DGLAP ln$k_t^2$ evolution. Moreover, the data already give hints that we need contributions not ordered in $k_t$, $\grave{a}~ la$ BFKL, since typically DGLAP overestimates the observed $\langle k_t \rangle$ and underestimates the mean multiplicity \cite{cms,atlas}. Another reason, both from a  theoretical and an  experimental viewpoint, is that it is not enough to have only one Pomeron ladder exchanged; we need to include multi-Pomeron exchanges.  
Theoretically, this ensures $s$-channel unitarity, and replaces the power 
growth of the cross section with energy by ln$s$ growth. First we have to 
include the unenhanced multi-Pomeron diagrams, which are usually described 
by a one- or two-channel eikonal model. Experimentally, 
these multi-Pomeron diagrams also explain the growth of the central 
plateau
\cite{cms,atlas}
\be
\frac{dN}{d\eta}~=~n_P \frac{dN_{\rm 1-Pom}}{d\eta},
\label{eq:nP}
\ee
where $dN_{\rm 1-Pom}/d\eta$ is the plateau due to the exchange of one 
Pomeron, which is independent of collider energy. The growth is due to the 
increasing number, $n_P$, of Pomerons exchanged as energy increases. These 
(eikonal) multi-Pomeron contributions are included in the present Monte 
Carlos to some extent, as a Multiple Interaction (MI)
option \cite{gmc}\footnote{In particular, the PYTHIA Monte Carlo uses a two-channel eikonal 
to calculate the probability of MI. However, low-mass dissociation of the 
incoming proton arising from the same two-channel eikonal is not included 
in the generator.}, but Pomeron-Pomeron interactions are not allowed for. 

In terms of partons, the main element of the formalism is the development of the parton cascade, which evolves mainly in ln$(1/x)$ space, and which is not strongly ordered in $k_t$. Nevertheless, the size of this parton cascade in impact parameter, $\vec b$, space is relatively small; that is, the $k_t$'s of the partons are not too low. Moreover, due to the BFKL diffusion in $\ln k_t$ space \cite{BFKL86}, the slope of the BFKL Pomeron trajectory $\alpha'_{\rm BFKL}\to 0$ (which is confirmed by the phenomenological analyses of high-energy data \cite{ost,KMRnn1,GLMM}). That is the transverse size of the cascade practically does not increase with energy.
In some sense, we may regard each cascade (that is, each Pomeron) as a small size `hot spot' inside both of the incoming protons\footnote
{Note that here we do not mean that a `hot spot' is simply a region inside one proton, but rather it is the parton cascade flowing from one proton to the other. We call this a hot spot since, in the relatively small domain in the $b$-plane occupied by the cascade, the density of partons is much larger than the average density over the whole interaction area. In fact, it is not unlike the situation simulated by the DGLAP-based Monte Carlos, where the cross section is given by the convolution of the matrix element of a hard central subprocess with the two parton cascades originating from the PDFs of the two colliding protons, see Fig.~\ref{fig:abc}(b). The only difference is that now we do not have strong $k_t$ ordering, and the structure of the matrix element which provides the matching of the two parton cascades does not differ qualitatively from the other parton interactions inside the cascades, see Fig.~\ref{fig:abc}(a). From this 
viewpoint a CCFM-based Monte Carlo, like CASCADE \cite{jung}, which uses 
 $k_t$-unintegrated parton densities and corresponding 
matrix elements, should be closer to reality.}. 
There may be several such hot spots and therefore high-energy $pp$ interactions mimic some features of nuclear-nuclear ($AA$) collisions.  That is, a high-energy $pp$ interaction mediated by the exchange of a few Pomerons is analogous to an $AA$ collision mediated by the interaction between a few pairs of constituent nucleons. In nuclear physics this is described by Glauber theory, whereas for a high-energy particle interaction it is known as the {\it eikonal} model, which accounts for the multiple rescattering of the incoming fast particles. 
Since the `hot spots' occur at 
different impact parameters, $b$, there is practically no interference 
between different chains (hot spots). Moreover, at this `eikonal' stage, 
the multi-Pomeron vertices, which account for the interaction between 
Pomerons, are not yet included in the formalism. We consider these interactions now.
\begin{figure} 
\begin{center}
\includegraphics[height=6.5cm]{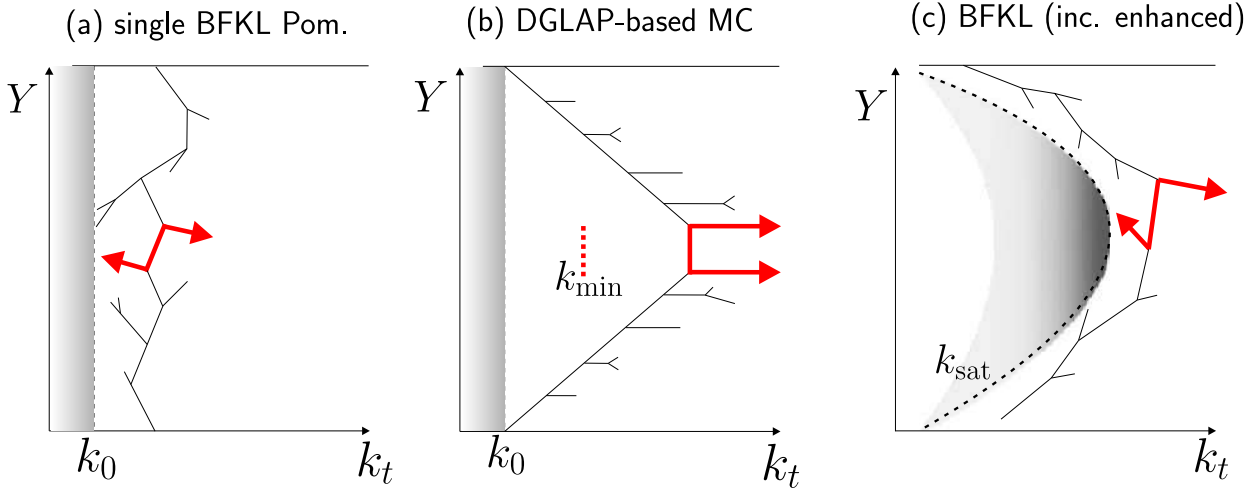}
\caption{\sf Schematic sketches of the basic diagram for semi-hard particle production in $pp$ collisions:  (a) BFKL-like 
 Pomeron exchange; (b) DGLAP-based Monte Carlos, where the hard matrix element has a cutoff $k_{\rm min}(s)$ tuned to the data; and (c), as in (a), but with the inclusion of enhanced multi-Pomeron diagrams (which suppresses parton production at low $k_t$, and hence effectively introduces a dynamically generated cutoff $k_{\rm sat}$ that depends on both the momentum fraction $x^+$ of the beam and $x^-$ of the target).}
\label{fig:abc}
\end{center}
\end{figure}

In addition to the above nucleon-nucleon interactions in an $AA$ collision, we also need to account for the interactions of secondaries, that are  produced in one nucleon-nucleon collision, with other nucleons in the colliding nuclei (or with other secondaries).  For $pp$ collisions this corresponds to interactions between partons within an individual hot spot (Pomeron). Formally, these are NNLO interactions, but their contribution is {\it enhanced} by the large multiplicity of partons within a high-energy cascade.  In terms of Reggeon Field Theory, the additional interactions are described by so-called {\it enhanced} multi-Pomeron diagrams, whose contributions are controlled by triple-Pomeron (and more complicated multi-Pomeron) couplings. Recall that non-enhanced (eikonal) multi-Pomeron interactions are caused mainly by Pomerons occurring at different impact parameters, and well separated from each other in the $b$-plane\footnote{Therefore the interaction between partons from two different hot spots has a small probability.}. On the other hand, the enhanced contributions mainly correspond to additional interactions (absorption) within an {\it individual} hot spot, but with the partons well separated in rapidity.

The main effect of the enhanced contribution is the absorption of low $k_t$ partons. Note that the probability of these additional interactions is proportional to $\sigma_{\rm abs} \sim 1/k_t^2$, and their main qualitative effect is to induce a splitting of low $k_t$ partons into a pair of partons each with lower $x$, but larger $k_t$ (see \cite{KMRnnn} and sect.6 of \cite{shuv}). Effectively this produces an infrared cut-off, $k_{\rm sat}$, on $k_t$, and partly restores a DGLAP-like $k_t$-ordering within the cascade at larger $k_t$.

\section{Schematic sketches of the model}

Qualitatively, the structure of soft interactions based on the `BFKL' multi-Pomeron approach is as follows. The evolution 
 produces a parton cascade which occupies a relatively small domain in $b$-space, as compared to the size of the proton. We have called this a hot spot. The multiplicity of partons grows as $x^{-\Delta}$, while the $k_t$'s of the partons are not strongly ordered and depend weakly on ln$s$. Allowing for the running of $\alpha_s$, the partons tend to drift to lower $k_t$ where the coupling is larger. This is shown schematically in Fig.~\ref{fig:abc}(a). 

On the contrary, the DGLAP-based Monte Carlos generate parton cascades strongly ordered in $k_t$. That is, the parton $k_t$ increases as we evolve from the input PDF of the proton to the matrix element of the hard subprocess, which occurs near the centre of the rapidity interval, Fig.~\ref{fig:abc}(b).  Since the cross section of the hard subprocess behaves as $d\hat{\sigma}/dk_t^2 \propto 1/k_t^4$, the dominant contributions come from near the lower limit $k_{\rm min}$, of the $k_t$ integration. In  fact, in order to describe the high-energy collider data, it is necessary to artificially introduce an energy dependent infrared cutoff; $k_{\rm min} \propto s^a$ with $a \sim 0.12$ \cite{P81}. This cutoff is only applied to the hard matrix element, whereas in the evolution of the parton cascade a constant cutoff $k_0$, corresponding to the input PDFs, is used. Note that during the DGLAP evolution, the position of the partons in $b$-space is frozen. Thus such a cascade also forms a hot spot.

Accounting for  the multiple interaction option, that is for contributions containing a few hot spots (that is, a few cascades), we include the eikonal multi-Pomeron contributions, both for the DGLAP and BFKL based descriptions.

Next, we include the enhanced multi-Pomeron diagrams 
introducing the absorption of the low $k_t$ partons. The strength of absorption is driven by the parton density and therefore the effect grows with energy, that is with ln$(1/x)$. We thus have an effective infrared cutoff, $k_{\rm sat}(x)$, which modifies the $k_t$ distribution of the `BFKL' cascade. The result is shown Fig.~\ref{fig:abc}(c), which has some similarity to the DGLAP cascade of Fig.~\ref{fig:abc}(b). However, now the cutoff $k_{\rm sat}$ is not a tuning parameter, but is generated dynamically by the enhanced multi-Pomeron diagrams. Recall that the same diagrams describe high-mass proton dissociation.  That is, the value of the multi-Pomeron vertex simultaneously controls the cross sections of high-mass dissociation and the effective cutoff $k_{\rm sat}$ -- two phenomena which, at first sight, appear to be quite different.

How may this be implemented in practice? In the model of Ref.~\cite{KMRnnn} the absorption of low $k_t$ partons is 
driven by the opacity, $\Omega$, which depends both on $k_t$ and 
$y=\ln(1/x)$. This opacity is obtained by solving the corresponding 
evolution equations in $y$. After the opacity is calculated, it can then be used in 
Monte Carlo generators to suppress the emission of low $k_t$ partons and, in 
this way, to introduce the effective cutoff $k_{\rm sat}$.

\section{Probes of `hard' properties of `soft' interactions at the LHC}

How do the properties described above reveal themselves in the LHC experiments?  The main ingredients of the framework, which we would like to confirm in semi-soft data at the LHC, are
\begin{itemize}
\item multi-Pomeron contributions arising from (unenhanced) eikonal diagrams; that is, the presence of hot spots;
\item multi-Pomeron contributions arising from enhanced diagrams (which describe the absorption of low $k_t$ partons and introduce an effective infrared cutoff, $k_{\rm sat}$, which increases with collider energy);
\item the presence of minijets. (Due to the  cutoff, $k_t>k_{\rm sat}$
the main inelastic process is minijet production.)
\end{itemize}

\subsection{Probe of hot spots and eikonal contributions   \label{sec:51}}

The presence of the eikonal multi-Pomeron contributions leads to the growth of the particle density, $dN/d\eta$, as energy increases. Recall that the central plateau arising from one-Pomeron exchange is independent of collider energy, and that the growth comes from the increase in the mean number of Pomerons, $n_P$, see (\ref{eq:nP}).

The presence of hot spots may be observed in Bose-Einstein Correlations (BEC) by selecting events with low particle density, that is, with $n_P \simeq 1$. Then the radius $r$ measured by BEC reflects the size of the domain from which a single hot spot emits secondaries. On the other hand events with large particle density are produced in collisions with several hot spots. Then two identical pions are likely to be emitted from different hot spots (different cascades) and the radius $r$ measured by BEC reflects their separation.  Therefore we expect $r$ to increase with increasing particle density
\cite{BEC}.

\subsection{Probe of enhanced diagrams via minijets and correlations}

The main effect induced by the enhanced multi-Pomeron diagrams is the absorption of low $k_t$ partons, leading to the growth of the dynamical infrared cutoff, $k_{\rm sat}$, with energy.  For high-energy `soft' $pp$ collisions the production of minijets with $p_t \gapproxeq k_{\rm sat}$ becomes the main source of secondaries.  However it is difficult to select an individual minijet, with such a relatively low $p_t$, in the high multiplicity LHC events. Nevertheless, the presence of minijets should reveal itself in two-particle correlations
\be
R_2(p_1,p_2)=\frac{d^2N/d^3p_1d^3p_2}{(dN/d^3p_1)\  (dN/d^3p_2)}-1
\ee
Without any correlations we would have a Poisson distribution with $R_2=0$.

Minijet (or resonance) production leads to $R_2>0$ when the rapidity
difference, $\delta \eta=|\eta_1-\eta_2|$ and the azimuthal angle $\phi$
between $\vec p_{1t}$ and $\vec p_{2t}$ are small.
It is informative \cite{rf} to select three different intervals of $\phi$ \\
$~~~~$(T) Toward$~~~$ -- $-\pi/4<\phi<\pi/4$ = the same jet\\
$~~~~$(B) Backward -- $3\pi/4<\phi<5\pi/4$ = backward jet\\
$~~~~$(A) Away$~~~~~$ -- $\pi/4<\phi<3\pi/4$ plus $5\pi/4<\phi<7\pi/4$ = away from jets,\\
and to compare the behaviour of $R_2$ in these 3 intervals. In the toward region both particles are likely to come from the same minijet.

The presence of minijets are expected to be revealed, by a stronger peak (of small width $\delta\eta \sim 1$)  in $R^T_2$, together with a weaker peak in $R^B_2$ (smeared out over a rather large $\delta\eta \sim 2-3$ interval), and almost nothing in 
$R^A_2$ (away region).
The effect should be stronger for the particles of relatively large $p_t\sim 1 -2$ GeV.
Such behaviour of $R_2^{T,B,A}$ will indicate the production of two groups of particles with relatively large $p_t$ which more or less balance each other. That is, a single minijet gives $R_2^T>0$, while the nearest minijet which balances the $p_t$ of the first gives $R_2^B>0$. In the case of resonance decay, we expect $R_2^T>0$, but small $R_2^B$.

It would also be interesting to    see how the correlation $R_2$ depends on $N_{\rm ch}$. In our approach there are two types of correlations: short-range correlations due to minijet production and long-range correlations \cite{shuv} due to events with several Pomerons (that is, several hot spots). By selecting events with large $N_{\rm ch}$ we select processes with many Pomerons. This leads to a decrease (or dilution) of the short-range correlation. It would be interesting to fix the number of Pomerons by measuring $N_{\rm ch}$ in one rapidity interval and to study the correlation in another interval. In the Pomeron approach we would expect to observe the same dilution of $R_2$.
  However if the large multiplicity is due to the decay of a `fireball' or other heavy object, then large $N_{\rm ch}$ should not affect the correlations observed in the domain separated from the object by a large rapidity interval.

\subsection{Probe of the multi-Pomeron vertex via high-mass dissociation}

It is important to study the interaction between the Pomerons explicitly. As we have emphasized the enhanced multi-Pomeron diagrams describe high-mass dissociation of the proton, as well as providing a dynamical infrared cutoff on $k_t$. Dissociation produces events with large rapidity gaps. These may be selected by vetoing the hadron activity in a
 large part of the available rapidity interval.
 Besides measuring the cross section of such a process, it would be valuable to study the spectrum (and the correlations) of secondaries near the edge of the rapidity gap. That is the secondaries in the Pomeron fragmentation region.

Since the triple-Regge analyses indicate that the transverse size of the triple-Pomeron vertex is very small 
\cite{luna}\footnote{For early references see, for instance, \cite{ABK,kktp}.},
this will lead to large $\langle p_t \rangle$ of the secondaries near the edge of the rapidity gap. An analogous phenomena was observed in deep inelastic scattering (DIS) where the $p_t$ of the secondaries with respect to the virtual {\it photon direction} are much larger in the current 
(point-like photon) fragmentation region than in the proton fragmentation region.

Contrary to single dissociation (SD), where the momentum transfer, $q_t$, across the gap is limited by the proton form factor, for double dissociation (DD) $q_t$ across the gap is controlled by the small size of the triple-Pomeron vertex, and hence may be rather large. (In DIS this $q_t$ is analogous to the momentum transfer
 of the virtual photon with respect to the {\it electron}.) 
The momentum $q_t$ will be carried by the secondaries produced near the 
edge of the rapidity gap, $\sum \vec{p}_t \simeq \vec{q}_t$, see Fig~\ref{fig:DD}. A large value of $\sum \vec{p}_t$ indicates DD with a large momentum transfer $q_t$. This would confirm the small size of the triple-Pomeron vertex.
\begin{figure} 
\begin{center}
\includegraphics[height=5cm]{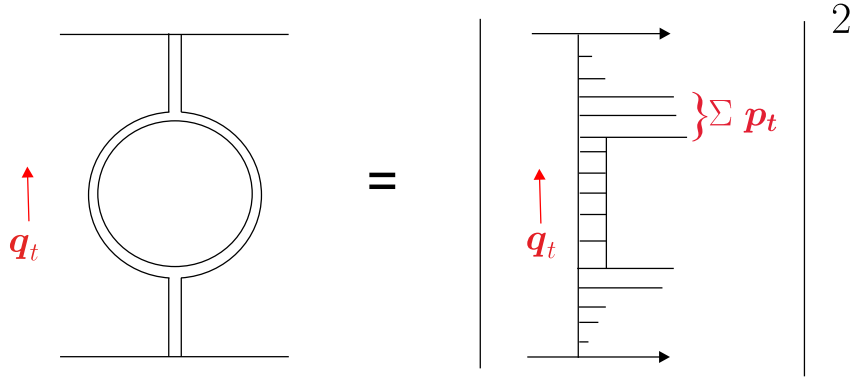}
\caption{\sf Secondaries produced near the edge of the rapidity gap in DD. A large value of $\sum \vec{p}_t$ would confirm the small size of the triple-Pomeron vertex.} 
\label{fig:DD}
\end{center}
\end{figure}


\subsection{Double-Pomeron-exchange processes}

An interesting possibility to study the Pomeron-Pomeron interaction is the so-called double-Pomeron-exchange (DPE) processes, where the centrally produced system is separated from the initial protons by large rapidity gaps.  It would be interesting to study the particle density, $dN/d\eta$,  the size of the interaction region by BEC, and the particle content of secondaries, in Pomeron-Pomeron collisions.  It is possible that in this case we will have a larger yield of $\eta,~\eta'$ mesons or even glueballs, that is gluon-rich particles, see section 7 of \cite{shuv} for more details. 
Since the indications are that the  size of the triple-Pomeron vertex is small, 
 we expect that the size of the Pomeron-Pomeron interaction region should be much less that the $pp$ interaction region. It means that the radius $r$ measured via BEC in DPE should be smaller than that discussed in Section \ref{sec:51}.

Recall that the  DPE cross sections, expected at the LHC, are not too small. For `elastic' recoil protons, observed in the forward proton detectors, it was estimated \cite{KMRprosp,shuv} to be of the order of 
$\xi_1\xi_2 d\sigma/d\xi_1 d\xi_2\sim 1\ -\ 5\,\mu b$. Here $\xi_i$ is the momentum fraction lost by proton $i$.  Integrating over
 the rapidity interval $\Delta\ln\xi= 2\ -\ 3$ we get $\sigma^{\rm DPE}\sim 10 - 30\ \mu b$. 
So the properties of these processes may be studied,
 by observing `elastic' protons in the forward proton detectors, 
already in the special  low luminosity  LHC runs, where the forward proton detectors will be used to measure the elastic $pp$ cross section in the low $t$ region. At least we have a chance to measure the density of secondaries
 $dN/d\eta$  produced in  Pomeron-Pomeron collisions, to measure the DPE cross section,   $\xi_1\xi_2 d\sigma/d\xi_1 d\xi_2$, and, may be, to determine the flavour of secondaries. In particular, we may check whether the yield of gluon-rich $\eta,\ \eta'$ mesons is enhanced in the DPE process.

\section{Ridge effect}
 Another manifestation of Pomeron-Pomeron interactions 
is the so-called `ridge' effect, that is the observation of ``Long-Range Near-Side Angular Correlations'' of secondaries produced in $pp$ \cite{cms2} or nuclear-nuclear \cite{AA} collisions. Many models have been proposed to explain this effect. As examples, we may mention 
the recent papers \cite{wer} and \cite{ven}, where one can find references to earlier studies. Actually in any model the effect is caused by
an additional interaction formulated either in terms of a hydrodynamic expansion
\cite{wer} or in terms of Pomeron ladders \cite{ven}.
However, for the `ridge' observed in $pp$ collisions at the LHC \cite{cms2}, the number of secondaries is not high enough to use thermodynamic variables. Here, we have only a few extra rescatterings.
So we present a simplified semi-quantitative estimate that demonstrates that the observed correlation
is consistent with an additional interaction of a relatively high-$E_T$ minijet.

Recall that the `ridge'-correlations are observed between the particles with
transverse momenta in the range $1 <p_t <3$ GeV in the events with large multiplicity
 $N_{\rm ch}>110$ \cite{cms2} (at  lower $p_t$ and/or $N_{\rm ch}$ the effect 
 is small). A minijet provides a relatively large $p_t$, while the large 
 multiplicity selects the contribution with a larger number of Pomerons.
The correlation presented by CMS is defined as
\be
R=(\langle N_{\rm ch} \rangle -1)\left(\frac{d^2N^{\rm signal}/d^2p_1d^2p_2}
{d^2N^{\rm mixed}/d^2p_1d^2p_2}-1\right).
\ee
For $N_{\rm ch}>110$  and $p_t=1-3$ GeV,  the value of $R$ reveals a peak at $\Delta\phi=0$ of about 
$\Delta R_{\rm CMS}\sim 1-2$ in height, in a wide region of $\Delta\eta$.

\begin{figure} 
\begin{center}
\includegraphics[height=7cm]{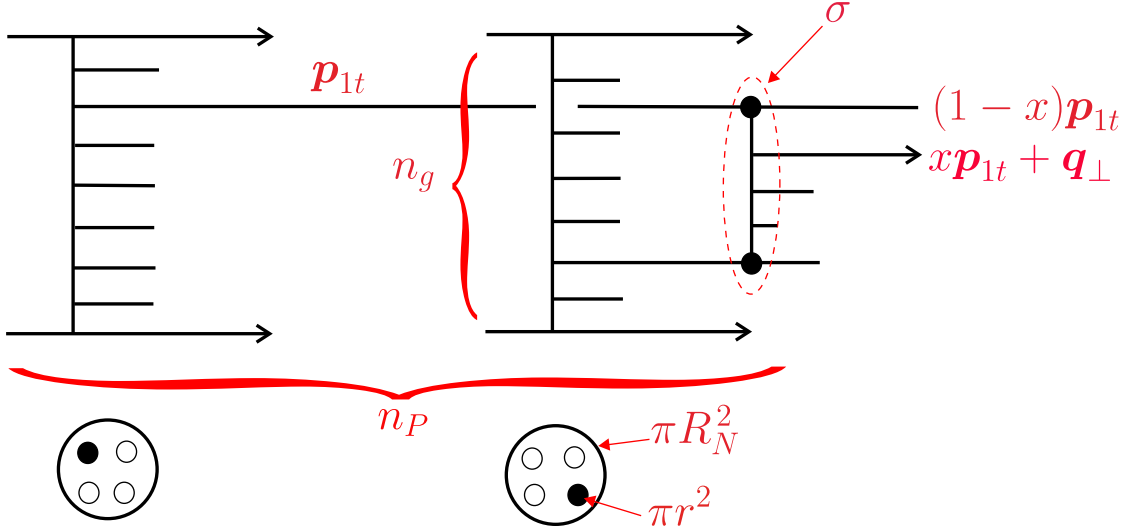}
\caption{\sf The rescattering of a minijet of transverse momentum $\vec{p}_{1t}$ producing a minijet of transverse momentum $x\vec{p}_{1t}+\vec{q}_\perp$ which is responsible for the observed ridge effect. The circles below the diagram show that the rescattering (with cross section $\sigma$) creates a cascade between two different hot spots (solid dots) in the proton. The probability of such a rescattering is enhanced by the number, $n_g$ of `target' gluons and the number $n_P$, of hot spots, see (\ref{eq:enh}).}
\label{fig:ridge}
\end{center}
\end{figure}

The origin of the peak is naturally explained by an additional inelastic interaction of a minijet. We denote the minijet by its  momentum $p_1$. It will produce a new set of secondaries, $q$. A secondary carries a fraction $x$ of the parent momentum $p_1$; that is, $\vec{q}_t=x\vec{p}_{1t}+\vec{q}_\perp$ where $\vec{q}_\perp$ is perpendicular to the direction of $\vec{p}_1$, see Fig. \ref{fig:ridge} \footnote{Fig. \ref{fig:ridge} shows the interaction of a parton from one hot spot with a parton from another hot spot. There may be a similar rescattering inside a single hot spot. The first is enhanced by $n_P$, while the second is enhanced due to the small area of the hot spot $\pi r^2 \ll \pi R^2_N$. However, the $p_t$ distribution originating from a single hot spot is strongly washed out by the uncertainty principle, $\Delta p_t \sim 1/r$. Therefore the correlation effect observed at large $N_{\rm ch}$ is mainly due to the structure indicated in Fig. \ref{fig:ridge}.}. In other words the new parton
 (minijet) adjacent to the parent parton will carry some part of its 
 original transverse momentum and therefore will preferentially go in the same 
azimuthal direction ($\Delta\phi\sim 0$). However, this is not the original minijet. Typically the two outgoing minijets are separated by a rapidity interval
$\Delta\eta\sim 2$. That is $\langle x\rangle \simeq e^{-\Delta\eta}\sim 0.1 - 0.2$ and the 
expected correlation
\be
R\sim (\langle N_{\rm ch}\rangle-1)\cdot x\cdot n_{\rm parton}\cdot w,
\label{eq:4}
\ee
where $(\langle N_{\rm ch} \rangle-1)$ is the normalization factor used by CMS. The momentum fraction $x$ accounts for the precision of alignment of the second minijet with respect to the parent direction. The remaining factor, $n_{\rm parton}w$, is the probability of rescattering and is given by the effective optical density, $n_{\rm parton}/\pi R^2_N$, multiplied by the cross $\sigma$ of the interaction. Here $R_N$ is the transverse radius of the proton.  Thus already we see that the rather small probability of a secondary interaction, $w\sim 1\%$,
is enhanced by the large normalization factor and the large number of appropriate targets, $n_{\rm parton}$; hence
 the value $\Delta R_{\rm CMS}\sim 1-2$ observed by CMS \cite{cms2} does not look surprising. 
A small correlation effect is considerably enhanced by the specific cuts that have been applied to the data.

Let us make a semi-quantitative estimate. By selecting  events with $N_{\rm ch}>110$, that is with about 4 times the
multiplicity of unbiased events, CMS choose processes
with a large number, $n_P$, of Pomerons (hot spots)\footnote{Note that in a usual minimum-bias high-energy (LHC) event we already observe a few Pomerons.} typically $n_P \sim 5-10$. Besides this, in each hot spot there are at least a few appropriate `target' gluons (say, $n_g\sim 3-5$). So, based on  model \cite{KMRnnn}, we get for the CMS cuts 
\be
n_{\rm parton}~=~n_P~ n_g\sim 10\ - 50.
\label{eq:enh}
\ee

To approximately estimate the probability, $w$, of the interaction with one parton we can use {\it either}  naive BFKL {\rm or} a pure `soft' approach since actually the transverse momenta, $p_t$, are not  large, $p_t \sim 1-3$ GeV. Recall that the probability 
\be
w=\sigma/\pi R^2_N.
\ee
It is driven, in the BFKL case, by the cross section 
\be
\sigma\sim \frac{ 4\pi\alpha^2_s}{\langle p^2_t \rangle }~ {\rm exp}(\Delta\delta y) \sim 0.7~{\rm mb},
\label{eq:s1}
\ee
where the exponential factor is the BFKL enhancement of the Low-Nussinov cross section; we take the resummed BFKL intercept $\Delta \simeq 0.3$. Also we have taken  $p_t=1.5$ GeV,  and $\delta y\sim 4$ to allow for rescattering on all the $n_g$ targets, rather than just the nearest one. On the other hand, for the `soft' estimate we use the known $\pi p$ cross section, $\sigma(\pi p)\sim 20$ mb, where the characteristic transverse momenta are of the order of $\langle k_t\rangle\sim 0.35$ GeV.
Taking  $p_t\sim 1.5$ GeV we expect a smaller cross section 
\be
\sigma~\sim ~(\langle k_t\rangle/\langle p_t\rangle)^2~20~\mbox{mb}~\sim~ 1 ~{\rm mb}.
\label{eq:s2}
\ee
 That is, both estimates give
$\sigma\sim 0.5 - 1$ mb, as compared to the transverse area of the proton  $\pi R^2_N\sim 20$ mb. Thus the probability of an interaction with one parton is about $w\sim 0.02- 0.05$.
Altogether this gives, using (\ref{eq:4}), a value  $\Delta R> 2$, which is 
compatible to that observed at the LHC by the CMS collaboration.

In  a larger $p_t$ domain the effect will be weaker since the cross sections in (\ref{eq:s1}) and (\ref{eq:s2})
both behave as $\sigma\propto 1/p^2_t$, and so the probability of an interaction decreases as $p_t$ increases. On 
 the other hand, for smaller $p_t$ the azimuthal distribution 
 is washed out by $\delta q_\perp \sim 0.5$ GeV coming from  hadronization;
in the limit of $p_t\to 0$ we have {\it no direction} at all. So, the correlation effect disappears.

Clearly in heavy ion ($AA$) collisions the effect should be larger due to the much larger parton densities. Neglecting the enhanced absorptive corrections we may expect, at a fixed impact parameter $b$, a parton density which is larger by up to a factor $A_1^{1/3}A_2^{1/3}$, that is by about a factor 35 for lead-lead interactions.

\section{Conclusion}
We can summarize the main features of high-energy soft and semihard interactions as follows.
\begin{itemize}
\item The interactions are mediated by parton cascades -- that is by Pomerons. The transverse size of a cascade is small in comparison with that of the proton.
\item Several parton cascades can occur in one collision -- the number of Pomerons increases with collider energy.
\item The interaction between Pomerons, that is between partons from
different cascades, leads to the absorption of low $k_t$ partons; thus we have an effective infrared cutoff, $k_{\rm sat}$, whose value grows with collider energy.
For this reason, minijet fragmentation is the main source of secondaries.
\item Most importantly, the dynamical multi-Pomeron generation of the cutoff, $k_{\rm sat}$ (and the weak dependence of the Pomeron intercept on scale) provides a natural smooth
transition from the `hard' to the  `soft' description of high-energy interactions.  We have only one sort of Pomeron -- the QCD BFKL-like Pomeron can be extended 
to describe soft interactions.
\end{itemize}
We have proposed possible measurements that can be made at the LHC which could illuminate the mechanism of multiparticle production. They can be summarized as follows.
\begin{itemize} 
\item To measure the size of the source of secondaries, which originate from one individual Pomeron (cascade), by studying identical pion Bose-Einstein correlations in low multiplicity events. (Recall low $N_{\rm ch}$ events originate from only one Pomeron exchange.)
\item To search for the two-particle correlations, $R_2$, caused by minijet production.
\item To study the spectra of secondaries near the edge of a Large Rapidity Gap (LRG) in order to probe the structure of the three-(or multi-)Pomeron vertex. 
\item A good possibility to study the Pomeron-Pomeron interaction directly is to select the Double-Pomeron-Exchange (DPE) events with a LRG on either side.
\end{itemize}
Clearly the Multiple Interactions (described, within a MI option in the present `general purpose' Monte Carlo generators, as a few independent parton cascades) are not actually independent from each other. In particular, this fact is manifested in the 
observation of the famous `ridge effect' at the LHC.  Indeed, we showed that additional minijet rescattering gives a natural semi-quantitative explanation of the effect.

The proposed measurements will improve our understanding of the dynamics
 of high-energy collisions, and additionally constrain 
 the models that are used to describe the diffractive and inelastic cross sections and the structure of underlying events. 

\section*{Acknowledgements}
MGR would like to thank the IPPP at the University of Durham for hospitality. This work was supported by the grant RFBR 11-02-00120-a
and by the Federal Program of the Russian State RSGSS-65751.2010.2.
This work is also supported in part by the network PITN-GA-2010-264564
(LHCPhenoNet).

{J. Adams, {\it et al.}}, [STAR Collaboration] Phys. Rev. {\bf C} {\bf
73}, 064907
  (2006);

A.~Adare {\it et al.}  [PHENIX Collaboration],
  Phys.\ Rev.\  C {\bf 78}, 014901 (2008);

  B.~I.~Abelev {\it et al.}  [STAR Collaboration],
  Phys.\ Rev.\  C {\bf 80}, 064912 (2009);

  B.~Alver {\it et al.}  [PHOBOS Collaboration],
  Phys.\ Rev.\ Lett.\  {\bf 104}, 062301 (2010);

  B.~Alver {\it et al.}  [PHOBOS Collaboration],
  Phys.\ Rev.\  C {\bf 81}, 034915 (2010).

\bibitem{wer} K. Werner, Iu. Karpenko and T. Pierog, Phys. Rev. Lett. {\bf 106}, 122004 (2011).
\bibitem{ven} A. Dumitru, K. Dusling, F. Gelis {\it et al.},
Phys. Lett. {\bf B697}, 21 (2011).

\end{document}